\documentclass[10pt]{IEEEtran}
\usepackage[dvips]{graphicx}
\newcommand{\qed}{\fbox{}}
\newcommand{\ve}[1]{ \mbox{\boldmath$#1$} }
\newcommand{\defeq}{\stackrel{\triangle}{=}}

\newtheorem{definition}{Definition}

\begin{document}

\title{Gradient Descent Bit Flipping Algorithms for Decoding LDPC Codes } 
\author{Tadashi Wadayama$^\dag$, Keisuke Nakamura$^\dag$, Masayuki Yagita$^\dag$, \\
Yuuki Funahashi$^{\dag\dag}$,   Shogo Usami$^{\dag\dag}$, Ichi Takumi$^\dag$}

 \maketitle
 
\begin{abstract}
A novel class of bit-flipping (BF) algorithms for decoding low-density parity-check (LDPC) codes is presented. The proposed algorithms, which are called {\em gradient descent bit flipping (GDBF) algorithms}, can be regarded as simplified gradient descent algorithms. Based on gradient descent formulation, the proposed algorithms are naturally derived from a simple non-linear objective function.

\end{abstract}

$\dag$: Nagoya Institute of Technology, $\dag\dag$: Meijo University.\\

\section{Introduction}
Bit-flipping (BF) algorithms for decoding 
low-density parity-check (LDPC) codes \cite{Gallager} have been investigated extensively
and many variants of BF algorithms 
such as weighted BF (WBF) \cite{WBF}, modified weighted BF (MWBF) \cite{MWBF}, 
and other variants \cite{IMWBF,originalRRWBF,RRWBF} 
have been proposed.
The first BF algorithm was developed by Gallager \cite{Gallager}. 
In a decoding process of Gallager's algorithm, some unreliable bits
(in a binary quantized received word)
corresponding to unsatisfied parity checks are flipped for 
each iteration. The successors of Gallager's BF algorithm inherits 
the basic strategy of Gallager's algorithm, namely, 
find  unreliable bits and then flip them. 
Although the bit error rate (BER) performance of the BF algorithm
is inferior to that of the sum-product algorithm or the min-sum algorithm, 
in general, the BF algorithm enables us to design a much simpler decoder, 
which is easier to implement. Thus, bridging the performance gap between BF decoding and BP decoding is an important technical challenge. 

In the present paper, a novel class of BF algorithms for decoding LDPC codes is presented. The proposed algorithm, which are called {\em gradient descent bit flipping (GDBF) algorithms}, can be regarded as bit-flipping gradient descent algorithms. The proposed algorithms are naturally derived from a simple gradient descent formulation. The behavior of the proposed algorithm can be explained from the viewpoint of the optimization of a non-linear objective function.

\section{Preliminaries}

\subsection{Notation}
Let $H$ be a binary  $m \times n$ parity check matrix, where $n > m \ge  1$.
The binary linear code $C$ is defined by
$
C \defeq \{\ve c \in F_2^n:  H \ve c = \ve 0 \}, 
$
where $F_2$ denotes the binary Galois field.
In the present paper, a vector is assumed to be a column vector.
For convention, we introduce the bipolar codes $\tilde C$ corresponding to $C$ as follows:
$
\tilde C \defeq \{(1-2c_1, 1-2c_2,\ldots, 1-2 c_n)  : \ve c \in C \}.
$
Namely, $\tilde C$, which is a subset of $\{+1, -1\}^n$,
 is obtained from $C$ by using binary $(0,1)$ to bipolar $(+1,-1)$ conversion.

The binary-input AWGN channel is assumed in the paper, 
which is defined by
$
\ve y = \ve c + \ve z
$
($\ve c \in \tilde C$).  The vector $\ve z = (z_1,\ldots, z_n)$ is a white Gaussian noise vector
where $z_j(j \in [1,n])$ is an i.i.d. Gaussian random variable with zero mean and variance $\sigma^2$.
The notation $[a,b]$ denotes the set of consecutive integers from $a$ to $b$.

Let $N(i)$ and $M(j) (i \in [1,m], j \in [1,n])$ be 
$N(i) \defeq \{j \in [1,n]: h_{ij} = 1   \}$ and $M(j) \defeq \{i \in [1,m]: h_{ij} = 1   \}$
where $h_{ij}$ is the $ij$-element of the parity check matrix $H$.
Using this notation, we can write the parity condition as:
$
\quad \prod_{j \in N(i)} x_j  = 1 (\forall i \in [1,m] )
$
which is equivalent to $(x_1,\ldots, x_n) \in \tilde C$.
The value $\prod_{j \in N(i)} x_j \in \{+1, -1\}$ is called the $i$-th {\em bipolar syndrome} of $\ve x$.

\subsection{Brief review on known BF algorithms}

A number of variants of BF algorithms have been developed. We can classify 
the BF algorithms into two-classes: single bit flipping (single BF) algorithms and 
multiple bits flipping (multi BF) algorithms. In the decoding process of the single BF algorithm, 
only one bit is flipped according to its bit flipping rule. On the other hand, the multi BF algorithm
allows multiple bit flipping per iteration in a decoding process.
In general, although the multi BF algorithm shows faster convergence than the single BF algorithm,
the multi BF algorithm suffers from the oscillation behavior of a decoder state, which is not easy to control. The framework of the single BF algorithms is summarized as follows:

\vspace{0.4cm}
\fbox{
\begin{minipage}{8cm}
\begin{description}
\item [\underline{Single BF algorithm}]
\item [Step 1] For $j \in [1,n]$, let
$
x_j := \mbox{sign}(y_j).
$
Let $\ve x \defeq (x_1,x_2,\ldots, x_n)$.
\item[Step 2] 
If the parity equation
$
\prod_{j \in N(i)} x_j  = +1
$
holds for all $i \in [1,m]$, output $\ve x$, and then exit.
\item[Step 3]  Find the flip position given by
\begin{equation}
\ell := \arg \min_{k \in [1,n]} \Delta_k(\ve x).
\end{equation}
and then flip the symbol:
$
x_\ell : = - x_\ell. 
$ The function $\Delta_k(\ve x)$ is called an {\em inversion function}.
\item[Step 4] If the number of iterations is 
under the maximum number of iterations $L_{max}$, 
return to Step 2; otherwise output $\ve x$ and exit. 
\end{description}
\end{minipage}
}
\vspace{0.4cm}

The function $\mbox{sign}(\cdot)$ is defined by
\begin{equation}
\mbox{sign}(x) \defeq
\left\{
\begin{array}{ll}
+1, & x \ge 0 \\
-1,  & x < 0. 
\end{array}
\right.
\end{equation}
In a decoding process of the single BF algorithm, 
hard decision decoding for a given $\ve y$ is first performed, and 
$\ve x$ is initialized to the hard decision result.
The minimum of the inversion function $\Delta_k(\ve x)$ for $k \in [1,n]$ is 
then found\footnote{When $\Delta_k(\ve x)$ is an integer-valued function, we need a tie-break rule to resolve a tie.}.
An inversion function $\Delta_k(\ve x)$ can be seen as a measure of the invalidness of bit assignment on $x_k$. 
The bit $x_\ell$,  where $\ell$  gives the smallest value of the inversion function,  is then flipped.

The inversion functions of WBF \cite{WBF} are defined by
\begin{equation}\label{invWBF}
\Delta_k^{(WBF)}(\ve x) \defeq \sum_{i \in M(k)} \beta_i \prod_{j \in N(i)} x_j.
\end{equation}
The values $\beta_i (i \in [1,m])$ is the {\em reliability} of 
bipolar syndromes defined by
$
\beta_i \defeq \min_{j \in N(i)} |y_j|.
$
In this case, the inversion function $\Delta_k^{(WBF)}(\ve x)$ gives the measure of
invalidness of symbol assignment on $x_k$, which is given by the sum of the weighted bipolar syndromes.

The inversion functions of MWBF \cite{MWBF} has a similar form of the inversion function 
of WBF but it contains a term corresponding to a received symbol.
The inversion function of MWBF is given by
\begin{equation} \label{invMWBF}
\Delta_k^{(MWBF)}(\ve x) \defeq  \alpha |y_k| + \sum_{i \in M(k)} \beta_i \prod_{j \in N(i)} x_j,
\end{equation}
where the parameter $\alpha$ is a positive real number.


\section{Gradient descent formulation}

\subsection{Objective function}
It seems natural to consider that
the dynamics of a BF algorithm as a minimization process of a hidden objective function.
This observation leads to a gradient descent formulation of BF algorithms.

The maximum likelihood (ML) decoding problem for the binary AWGN channel is equivalent to 
the problem of finding a (bipolar) codeword in $\tilde C$, which gives 
the largest correlation to a given received word $\ve y$. Namely, the MLD rule 
can be written as 
$
\hat{\ve x }
= \arg \max_{\ve x \in \tilde C} \sum_{j=1}^n x_i y_i.
$

Based on this correlation decoding rule, 
we here define the following objective function:
\begin{equation}\label{GDBFobjective}
f(\ve x) \defeq \sum_{i=1}^n x_j y_j + \sum_{i=1}^m \prod_{j \in N(i)} x_j.
\end{equation}
The first term of the objective function corresponds to the correlation between a bipolar codeword and 
the received word, which should be maximized. The second term is the sum of the bipolar syndromes of $\ve x$.
If and only if $\ve x \in \tilde C$, then the second term has its maximum value 
$
\sum_{i=1}^m \prod_{j \in N(i)} x_j = m.
$
Thus, this term can be considered as a {\em penalty term}, which forces $\ve x$ to be a valid codeword.
Note that this objective function is a non-linear function and has many local maxima. These 
local maxima become a major source of sub-optimality of the GDBF algorithm presented later.

\subsection{Gradient descent BF algorithm}

For the numerical optimization problem for a differentiable function such as (\ref{GDBFobjective}), 
the gradient descent method \cite{Boyd} is a natural choice for the first attempt.
The partial derivative of $f(\ve x)$ with respect to the variable $x_k (k \in [1,n])$ can be immediately derived from the definition of $f(\ve x)$: 
\begin{equation}
\frac{\partial}{\partial x_k}f(\ve x) = y_k  + \sum_{i \in M(k)} \prod_{j \in N(i) \backslash k } x_j.
\end{equation}

Let  us consider  the product of $x_k$ and the partial derivative of $x_k$ in $\ve x$, namely
\begin{equation}
x_k \frac{\partial}{\partial x_k}f(\ve x) = x_k y_k  + \sum_{i \in M(k)}  \prod_{j \in N(i)} x_j.
\end{equation}
For a small real number $s$, we have the first-order approximation:
\begin{equation} 
f(x_1, \ldots, x_k+ s,\ldots, x_n )  \simeq f(\ve x) + s  \frac{\partial}{\partial x_k}f(\ve x).
\end{equation}
When $ \frac{\partial}{\partial x_k}f(\ve x) > 0$, we need to choose $s  >  0$
in order to have 
\begin{equation}\label{larger}
f(x_1, \ldots, x_k+ s,\ldots, x_n )  >  f(\ve x).
\end{equation}
On the other hand, if $ \frac{\partial}{\partial x_k}f(\ve x) < 0$ holds,  we should choose $s < 0$ 
to obtain the inequality (\ref{larger}).
Therefore, if $x_k \frac{\partial}{\partial x_k}f(\ve x)  < 0$, then flipping the $k$th symbol ($x_k := -x_k$)
may increase the objective function value\footnote{ 
There is a possibility that the objective function value may decrease
because the step size is fixed (such as single flip).}.

One reasonable way to find a flipping position is 
to choose the position at which the absolute value of the partial derivative is largest.
This flipping rule is closely related to the steepest descent algorithm based on 
$\ell_1$-norm (also known as the {\em coordinate descent algorithm}) \cite{Boyd}.
According to this observation, we have the following rule to choose the 
flipping position.
\begin{definition}[Inversion function of the GDBF algorithm]
The single BF algorithm based on the inversion function
\begin{equation}\label{eqconv}
\Delta_k^{(GD)}(\ve x) \defeq   x_k y_k  + \sum_{i \in M(k)}  \prod_{j \in N(i)} x_j
\end{equation}
is called the Gradient descent BF (GDBF) algorithm. \hfill\qed
\end{definition}
Thus, the decoding process of the GDBF algorithm can be seen as the minimization process of 
$-f(\ve x)$ (it can be considered as the {\em energy} of the system) based {\em bit-flipping} gradient descent method.

It is interesting to see that the combination of the objective function $\tilde f(\ve x)$ defined by
\begin{equation}\label{wbfobj}
\tilde f(\ve x) \defeq \alpha \sum_{i=1}^n x_j |y_j| + \sum_{i=1}^m \beta_i \prod_{j \in N(i)} x_j
\end{equation}
and the argument on gradient descent presented above
gives the inversion functions of conventional algorithms such as the WBF algorithm (\ref{invWBF}) and the MWBF algorithm (\ref{invMWBF}).
However, this objective function (\ref{wbfobj}) looks less meaningful 
compared with the objective function (\ref{GDBFobjective}). 
In other words, the inversion function $\Delta_k^{(GD)}(\ve x)$ defined in (\ref{eqconv}) 
has a more natural interpretation than those of the conventional algorithms: $\Delta_k^{(WBF)}(\ve x)$ in (\ref{invWBF}) and $\Delta_k^{(MWBF)}(\ve x)$ in  (\ref{invMWBF}). Actually, the new inversion function $\Delta_k^{(GD)}(\ve x)$  is not only natural but also {\em effective} in terms of bit error performance and convergence speed.

\subsection{Multi GDBF algorithm}

A decoding process of  the GDBF algorithm can be regarded as a
maximization process of the objective function (\ref{GDBFobjective}) in a gradient ascent manner. Thus, we can utilize the objective function value in order to observe the convergence behavior.
For example, it is possible to monitor the value of the objective function for each iteration. 
In the first several iterations, the value increases as the number of iterations increases. 
However, the value eventually ceases to increase when 
the search point arrives at the nearest point in $\{+1,-1\}^n$ to the local maximum 
of the objective function. We can easily detect 
such convergence to a local maximum by observing the value of the objective function.

\begin{figure}[htbp]
\begin{center}
\includegraphics[scale=0.4]{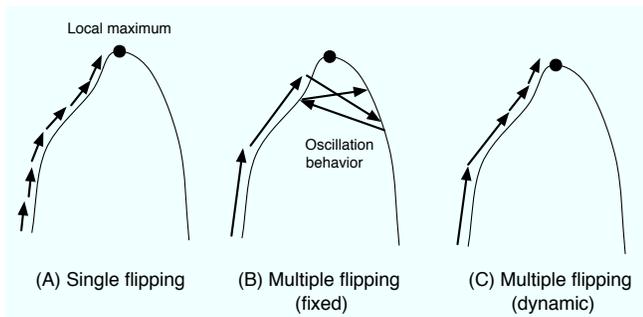} \\
(A) converging but slow, (B) not converging but fast, (C) converging and fast
  \caption{Convergence behavior}
  \label{localmax}
\end{center}
\end{figure}

Both the BF algorithms reviewed in the previous section and the GDBF algorithm 
flip only one bit for each iteration. In terms of the numerical optimization, 
in these algorithms, a search point moves towards a local maximum with a very small step (i.e., 1 bit flip) in order to avoid oscillation around the local maximum (See Fig.\ref{localmax} (A)). However, the small size step leads to slower convergence to a local maximum. In general, compared with the min-sum algorithm, BF algorithms (single flip/iteration) require a larger number of iterations to achieve the same bit error probability.

The multi bit flipping algorithm is expected to have a faster convergence speed than that of the single bit flipping algorithm because of its larger step size. If the search point is close to a local maximum, 
a fixed large step is not suitable to find the (near) local maximum point; it leads 
to oscillation behavior of a multi-bit flipping BF algorithm (Fig.\ref{localmax}(B)).
We need to adjust the step size dynamically from a large step size to a small step size 
in an optimization process (Fig.\ref{localmax}(C)). 

The objective function is a useful guideline for adjusting the step size (i.e., number of flipping bits). The {\em multi GDBF algorithm} is a GDBF algorithm including the multi-bit flipping idea. 
In the following, we assume the inversion function $\Delta_k^{(GD)}(\ve x)$ 
defined by (\ref{eqconv}) (the inversion function for the GDBF algorithm).

The flow of the multi GDBF algorithm is almost the same as that of the previously presented GDBF algorithm. When it is necessary to clearly distinguish two decoding algorithms, the GDBF algorithm presented in the previous sub-subsection is referred to as the single GDBF algorithm.  

In order to define the multi GDBF algorithm, we need to introduce new parameters $\theta$ and $\mu$. The parameter $\theta$ is a negative real number, which is called the {\em inversion threshold}. The binary (0 or 1) variable $\mu$, which is called the {\em mode flag}, is set to 0 at the beginning of the decoding process. Step 3 of the BF algorithm should be replaced with the following multi-bit flipping procedure.

\vspace{0.4cm}
\fbox{
\begin{minipage}{8cm}
\begin{description}
\item[Step 3] Evaluate the value of the objective function, and let $f_1:= f(\ve x)$.
If $\mu = 0$, then execute Sub-step 3-1 (multi-bit mode), else execute Sub-step 3-2 (single-bit mode).
\begin{description} 
\item[3-1 ] Flip all the bits satisfying 
\[
\Delta_k^{(GD)} < \theta\quad (k \in [1,n]).
\]
Evaluate the value of the 
objective function again, and let $f_2 := f(\ve x)$. If $f_1 > f_2$ holds, then let $\mu = 1$.
\item[3-2] Flip a single bit at the $j$th position, where
\[
j \defeq \arg \min_{k \in [1,n]} \Delta_k^{(GD)}.
\]
\end{description}
\end{description}
\end{minipage}
}
\vspace{0.4cm}

Usually, at the beginning of a decoding process, the objective function value increases as the number of iterations increases in the multi-bit mode, namely, $f_1 < f_2$ holds for the first few iterations. When the search point eventually arrives at the point satisfying $f_1 > f_2$, the bit flipping mode is changed from the multi-bit mode ($\mu = 0$) to the single-bit mode ($\mu = 1$). This mode change means adjustment of the step size, which helps a search point to converge to a local maximum when the search point is located close to the local maximum.

\section{Behavior of the GDBF algorithms}
In this section, the behavior and decoding performance of (single and multi) GF-BF algorithms 
obtained from computer simulations are presented. 

\begin{figure}[htbp]
\begin{center}
\includegraphics[scale=0.7]{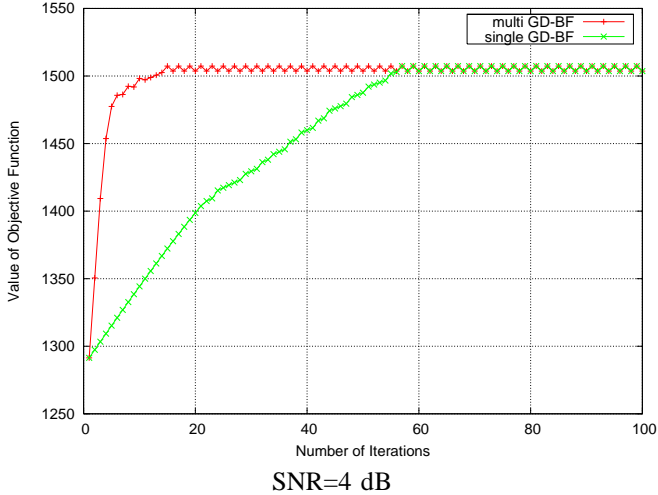} \\
SNR=4 dB
  \caption{Objective function values in GDBF processes as a function of the number of iterations}
  \label{regular-objval}
\end{center}
\end{figure}

Figure \ref{regular-objval} presents objective function values (\ref{GDBFobjective}) as a function of the number of iterations in the single and multi GDBF processes.
Throughout the present paper, a regular LDPC code with $n = 1008, m = 504$ 
(called PEGReg504x1008 in \cite{MacKayEn}) is assumed. The column weight of the code is 3.
In both cases (single and multi), we tested the same noise instance, and both algorithms
output the correct codeword (i.e., successful decoding).

In the case of the single GDBF-algorithm, 
the objective function value gradually increases as the number of iterations grows in the first
50--60 iterations. After the slope, the increment of the objective function value eventually stops, and a flat part that corresponds to a local maximum appears.
In the flat part of the curves, the oscillation behavior of the objective function value can be seen.
Due to the constraint such that a search point $\ve x$ must lie in $\{+1,-1\}$, a GDBF process 
cannot find a true local maximum point (the point where the gradient of the
objective function becomes a zero vector) of the objective function. Thus, a search point 
moves around the local maximum point. This move causes the oscillation behavior 
observed in a single GDBF process.
The curve corresponding to the multi GDBF algorithm shows much faster convergence compared 
with the single GDBF algorithm. It takes only 15 iterations for 
the search point to come very close to the local maximum point.

\begin{figure}[htbp]
\begin{center}
\includegraphics[scale=0.7]{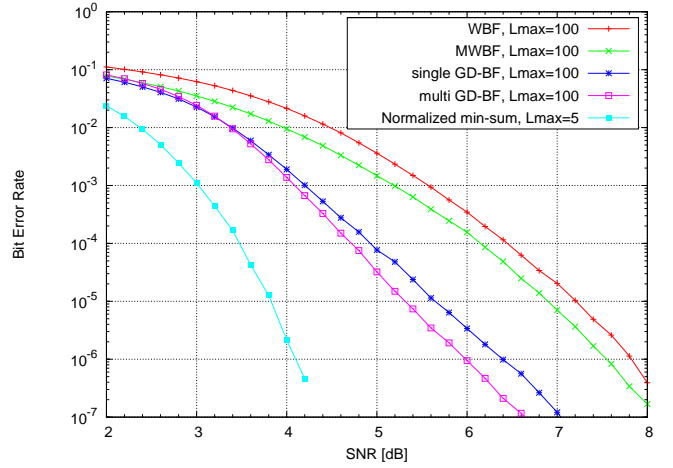} \\
  \caption{Bit error rate  of GDBF algorithms: regular LDPC code (PEGReg504x1008\cite{MacKayEn})}
  \label{regular-single}
\end{center}
\end{figure}

Figure \ref{regular-single} presents the bit error curves of single and multi GDBF algorithms ($L_{max} = 100, \theta=-0.6$). As references, the curves for the WBF algorithm ($L_{max} = 100$),  the MWBF algorithms ($L_{max} = 100, \alpha=0.2$), and the normalized min-sum algorithm ($L_{max} = 5$, scale factor 0.8) are included as well. The parameter $L_{max}$ denotes the maximum number of iterations for each algorithm. We can see that the GDBF algorithms perform much better than the WBF and MWBF algorithms. For example, at BER $= 10^{-6}$, the multi GDBF algorithm offers a gain of approximately 1.6 dB compared with the MWBF algorithm. Compared with the single GDBF algorithm, the multi GDBF algorithm has a steeper slope in its error curve. Unfortunately, there is still a large performance gap between the error curves of the normalized min-sum algorithm and the GDBF algorithms. The GDBF algorithm fails to decode when a search point is attracted to an undesirable local maximum of the objective function. This large performance gap suggests the existence of some local maxima relatively close to a bipolar codeword, which degrades the BER performance.

\begin{figure}[htbp]
\begin{center}
\includegraphics[scale=0.7]{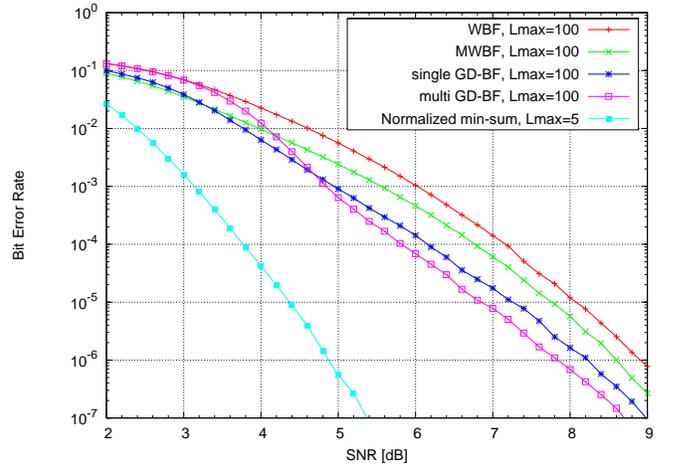} \\
  \caption{Bit error rate  of GDBF algorithms: irregular LDPC code (PEGirReg504x1008\cite{MacKayEn})}
  \label{irregular-single}
\end{center}
\end{figure}

Figure \ref{irregular-single} shows error curves for an irregular LDPC code.
The code used in the simulation is an irregular code (called PEGirReg504x1008 in \cite{MacKayEn}) 
constructed based on PEG construction. The same decoding algorithms (with same parameter)
appeared in Fig.\ref{regular-single} have been tested. As well as the regular case, the error curves of 
GD-BF algorithms come bellow those of WBF and MWBF algorithms. However, the
improvement is relatively small compared with the regular case. This observation may imply
that the advantage of GD-BF algorithm in BER depends on type of the code. 

\begin{figure}[htbp]
\begin{center}
\includegraphics[scale=0.7]{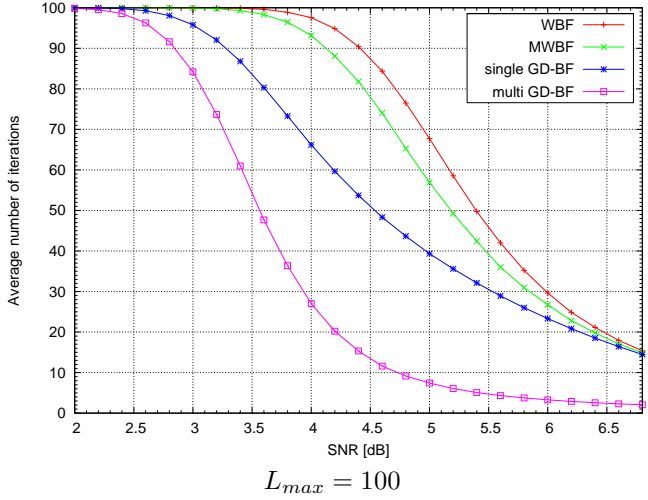} \\
$L_{max} = 100$
  \caption{Average number of iterations}
  \label{regular-aveitr}
\end{center}
\end{figure}

In order to evaluate the convergence speed of BF algorithms, 
the average number of iterations is an appropriate measure.
Figure \ref{regular-aveitr} shows the average number of iterations (as a function of SNR)
of the GDBF algorithms (single and multi), the WBF algorithm, and the MWBF algorithms.
Note that the multi GDBF algorithm certainly have a fast convergence property.
Large gaps can be observed between the curve of the multi GDBF algorithm and the other curves.

\section{Escape from a local maximum}

\subsection{Effect of non-codeword local maxima}

\begin{figure}[htbp]
\begin{center}
\includegraphics[scale=0.7]{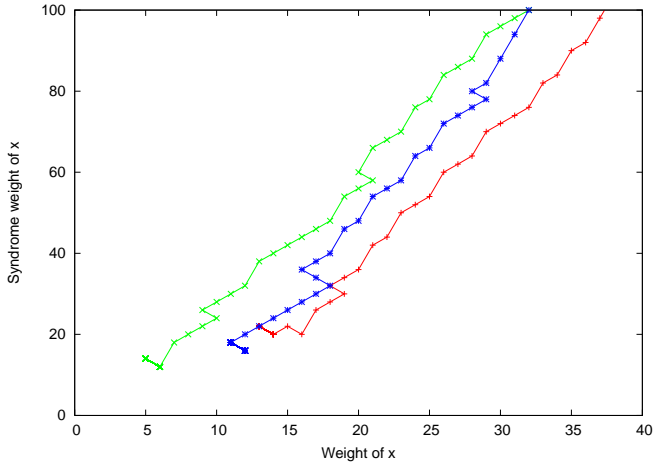} \\
  \caption{Trajectories of weight and syndrome weight of search points}
  \label{trajectory}
\end{center}
\end{figure}

As we have discussed, a decoding failure occurs when a search point is captured by a local maximum, which is not a transmitted codeword. Thus, it is desirable to know the effect of such local maxima. Figure \ref{trajectory} presents three trajectories of weight and syndrome weight of a search point in three decoding processes corresponding to decoding failure.
The weight of a search point $\ve x$ is defined by
$
w_1(\ve x) \defeq |\{j \in [1,n]: x_j = -1 \}|. 
$
In a similar way, the syndrome weight of $\ve x$ is given by
$
w_2(\ve x) \defeq \left| \left\{i \in [1,m]:  \prod_{j \in N(i)} x_j = -1 \right\} \right|.
$
We assume that the all-1 bipolar codeword (i.e., all-zero binary codeword) is transmitted 
without loss of generality.

We can obtain the following observation from Fig.\ref{trajectory}:
(i) the decoding process starts from the position at which both $w_1(\ve x)$ and $w_2(\ve x)$ are large, (ii) $w_1(\ve x)$ and $w_2(\ve x)$ decreases as the iteration proceeds, and (iii) the final states of the search point have a relatively small value of $w_1(\ve x)$ and $w_2(\ve x)$.

Based on these observations, we may be able to conjecture that 
a search point is finally trapped by a local maximum close to a {\em near codeword} 
in high probability\footnote{Note that other experiments also support this conjecture. }.
Near codewords \cite{Nearcodeword} are bipolar codewords of $\tilde C$ 
that have both small weight and small syndrome weight.
The sub-optimality of BF-algorithms compared with sum-product and min-sum algorithms
comes from the effect of these numerous local maxima.

\subsection{GDBF algorithm with escape process}
Since the weight of the final position of a search point is so small, a small perturbation of a captured search point appears to be helpful for the search point to escape from an undesirable local maximum. We can expect that such a perturbation process improves the BER performance of BF algorithms.

One of the simplest ways to add a perturbation on a trapped search point is
to switch the flip mode from the single-bit mode to the multi-bit mode with an appropriate threshold forcibly when the search point arrives at a non-codeword local maximum.
This additional process is called the {\em escape process}.
In general, the escape process reduces the object function value, i.e., 
the search point moves downwards in the energy landscape. After the escape process, the search point again begins to climb a hill, which may be different from the trapped point.

\begin{figure}[htbp]
\begin{center}
\includegraphics[scale=0.5]{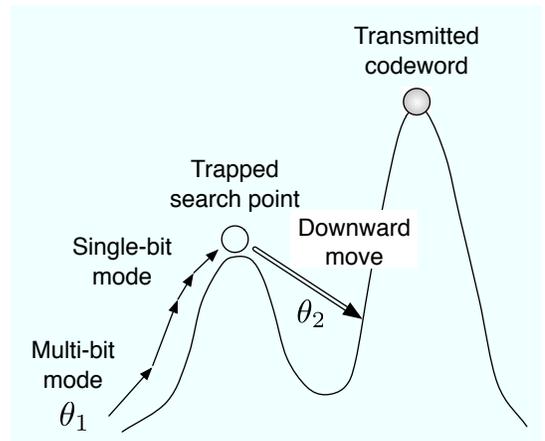} \\
    \caption{Idea of escape process}
  \label{escapeprocess}
\end{center}
\end{figure}

We here modify the multi GDBF algorithm by incorporating two thresholds: $\theta_1$ and  $\theta_2$. The threshold $\theta_1$ is the threshold constant used in the multi-bit mode at the 
beginning of the decoding process. After several iterations, the multi-bit mode
is changed to single-bit mode and then the search point may eventually arrive at the
non-codeword local maximum. In such a case, the decoder changes its mode to the multi-bit mode (i.e., $\mu = 0$) with threshold $\theta_2$. Thus, the threshold $\theta_2$ can be regarded as the threshold for {\em downward movement}.
Although $\theta_2$ can be a constant value, in terms of the BER performance, 
it is advantageous to choose randomly\footnote{This fact is observed from some experiments.}. 
In other words, $\theta_2$ can be a random variable.
After the downward move (just one iteration), the decoder changes the threshold to $\theta_1$ again. The above process continues until the parity check condition holds or the number of iterations becomes $L_{max}$.
Figure \ref{escapeprocess} illustrates the idea of the escape process.

\subsection{Simulation results}

\begin{figure}[htbp]
\begin{center}
\includegraphics[scale=0.7]{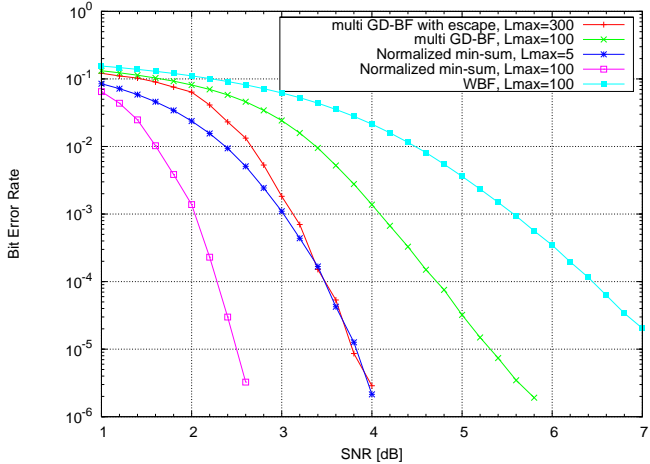} \\
    \caption{Bit error rate of the GDBF algorithm with the escape process}
  \label{escape}
\end{center}
\end{figure}

Figure \ref{escape} shows the BER curve of such a decoding algorithm (labeled 'multi GDBF with escape'). 
In this simulation, we used the parameters: $\theta_1 = -0.7, \theta_2 = 1.7 + \alpha$ where
$\alpha$ is a Gaussian random number with mean zero and variance 0.01.
These parameters have been obtained an ad hoc optimization at SNR = 4dB.
We can see that the BER curve of multi GDBF with escape (with $L_{max}=300$) is much steeper than that of the naive multi GDBF algorithm. At BER = $10^{-5}$, multi GDBF with escape achieves a gain of almost 1.5 dB compared with the naive multi GDBF algorithm. The average number of iterations of multi GDBF with escape is approximately 25.6 at SNR = 4 dB. This result implies that the perturbation can actually save some trapped search points to converge to the desirable local maximum corresponding to the transmitted codeword. It is an interesting open problem to optimize the flipping schedule to narrow the gap between the min-sum BER curve and the GDBF BER curve.

\section{Conclusion}
This paper presents a class of BF algorithms based on the gradient descent algorithm.
GDBF algorithms can be regarded as a maximization process of the object function 
using bit-flipping gradient descent method (i.e., bit-flipping dynamics which minimizes the energy $-f(\ve x)$).  
The gradient descent formulation naturally introduces 
an energy landscape of the state-space of the BF-decoder.
The viewpoint obtained by this formulation
brings us a new way to understand convergence behaviors of BF algorithms.
Furthermore this viewpoint  is also useful to design improved decoding algorithms
such as the multi GDBF algorithm and the GDBF algorithm with escape process from an undesired local maximum.
The GDBF algorithm with escape process performs very well compared with known BF algorithms.
One lesson we have learned from this result is that 
fine control on flipping schedule is indispensable to improve decoding performance of BF algorithms.

\subsection*{Acknowledgement}
The present study was supported in part by the Ministry of Education, Science, Sports,
and Culture of Japan through a Grant-in-Aid for Scientific Research on Priority Areas
(Deepening and Expansion of Statistical Informatics) 180790091
and by a research grant from the Storage Research Consortium (SRC).

\end{document}